\documentclass[lettersize,journal]{IEEEtran}
\usepackage{amsmath,amsfonts}
\usepackage{algorithmic}
\usepackage{array}
\usepackage[caption=false,font=normalsize,labelfont=sf,textfont=sf]{subfig}
\usepackage{textcomp}
\usepackage{stfloats}
\usepackage{url}
\usepackage{verbatim}
\usepackage{graphicx}
\usepackage{cite}
\usepackage{amssymb} % 一些特殊的字体（如\mathbb等）需要这个宏.
\usepackage{MnSymbol}
\usepackage{extarrows}% 长等号上下标注,  长箭头上下标注
\usepackage{enumerate} % 用来调整enumerate环境,  至少想要修改枚举的编号需要使用这个宏
\usepackage{bm} % 加粗字体包
\usepackage{booktabs}
\usepackage{algorithmic}
\usepackage{algorithm}

\newtheorem{theorem}{Theorem}[section]
\newtheorem{definition}{Definition}[section]

\newtheorem{corollary}{Corollary}[section]
\newtheorem{remark}{Remark}[section]
\renewcommand{\P}{\mathbb{P}}
\newcommand{\E}{\mathbb{E}}

\newcommand{\pf}{\textbf{Proof: }}
\newcommand{\e}{\hfill$\blacksquare$}

\usepackage{epstopdf}
\usepackage{color}
\newcommand{\tr}{\textcolor{black} }

\allowdisplaybreaks[4] % 允许公式跨页,  [a]表示跨页力度,  a越大,  跨页力度越大

\def\BibTeX{{\rm B\kern-.05em{\sc i\kern-.025em b}\kern-.08em
		T\kern-.1667em\lower.7ex\hbox{E}\kern-.125emX}}
\usepackage{balance}
\begin{document}
	\title{Erlang Model for \tr{Multi-type Data  Flow}}
	\author{Liuquan Yao,
		Pei Yang,
		Zhichao Liu,
		Wenyan Li,
		Jianghua Liu,
		and Zhi-Ming Ma
		\thanks{This work is supported by National Key R\&D Program of China No. 2023YFA1009601 and 2023YFA1009602.

			Liuquan Yao, Zhichao Liu and Zhi-Ming Ma are with
			the University of Chinese Academy of Sciences, Academy of Mathematics and Systems Science, CAS, Beijing 100190, China (e-mail: yaoliuquan20@mails.ucas.ac.cn, liuzhichao20@mails.ucas.ac.cn, mazm@amt.ac.cn).
			
			Pei Yang, Wenyan Li and Jianghua Liu are with
			Huawei Technologies Co. Ltd., China (e-mail: yangpei30@hisilicon.com, liwenyan1@huawei.com,  Liu.Jianghua@huawei.com).
	}}
	
	%\markboth{Journal of \LaTeX\ Class Files,~Vol.~18, No.~9, September~2020}%
	%{How to Use the IEEEtran \LaTeX \ Templates}
	
	\maketitle

	\begin{abstract}
		With the development of information technology, requirements for data flow have become diverse. When \tr{multi-type data flow (MDF)} is used, \tr{games, videos, calls, \textit{etc.} are all requirements. There may be a constant switch between these requirements, and also multiple requirements at the same time.} Therefore, the demands of users change over time, which makes traditional teletraffic analysis not directly applicable. This paper proposes probabilistic models for the requirement of MDF, and analyzes in three states: non-tolerance, tolerance and delay. \tr{When the requirement random variables are co-distributed with respect to time}, we  prove the practicability of the Erlang Multirate Loss Model (EMLM) from a mathematical perspective by discretizing time and error analysis. An algorithm of pre-allocating  resources  is given to guild the construction of base resources.
	\end{abstract}
	\begin{IEEEkeywords}
		Erlang Formula; \tr{ Multi-type Data Flow};  Poisson Process; Negative Exponential Distribution
	\end{IEEEkeywords}

	\section{Introduction}\label{section1}
	
	Communication has become an indispensable part of modern society. For a community, a large number of users will make communication requirements at the same time, and each requirement needs to allocate communication resources, such as telephone lines, time-frequency resource grids, etc. When infrastructure construction is carried out (such as base stations), if there are few preset communication resources, the user demand in the area will be frequently blocked, resulting in a poor user experience, while too many preset resources  will lead to increased costs and waste. Therefore, predicting the performance of users' requirements in the communication society and selecting reasonable resource presets are  important steps in infrastructure construction\cite{ONOFF-Finite}.
	
	In 1917, A.K. Erlang  obtained his famous formula  from the analysis of the statistical equilibrium and laid the foundations of modern teletraffic theory\cite{Ramalhoto2011}. By modeling the number of arrival users as Poisson random variable and the required time being exponential distributed, Erlang formula can deduce the blocking probability for telephone communication, according to the birth and death process theory.
	
	The original Erlang model was only for telephone line, and
	in order to be suitable with complex situations, Erlang formula has been sustainably developing. \cite{1965G} considered two types of requirements, narrow-band and wide-band, and calculate the related blocking probability, in 1965. With more analysis, the number of types can be generated to any integer $K$ and the model was called as \textit{Erlang Multirate Loss Model} (EMLM) \cite{BeyondEr}. Some specific policies on link, waiting or tolerance  were also added in the model to meet several situations \cite{BeyondEr}, \cite{tolerance2012},	\cite{THandTolerance}.  There are also some works on the assumptions of the arrival user and required time.  	\cite{finiteN} studied the case when the size of community is not large enough, and used quasi-random call arrival process to replace Poisson arrival process. In addition, the loss was thought as the noise when the band-demand exceeded the total band.  	\cite{ON-OFF} set two stages for activated users: \textit{ON} and \textit{OFF}, which can   summarize the demand characteristics of  games. 
	%When an user is \textit{ON}, he requires $b_k$ band if he is in class $k$, and when he finishes the stage ON, he becomes \textit{OFF} with probability $\sigma_k$ and leaves with probability $1-\sigma_k$. When an user finishes his \textit{OFF} stage, he ask to become \textit{ON}. This assumption is called as ON-OFF model, which was enhanced for finite population community \cite{ONOFF-Finite}. 
	\tr{Recently, \cite{M2M} made a more practical analysis by discussing machine-to-machine traffic model rather than human-to-human, and setting the arrival distribution as a Beta distribution over a period. 
		An Erlang model with varied cost in time was studied in \cite{optimal}, and \cite{no-mark} consider a single-server system without Markov property.}
	
	However,  with the development of wireless communication, modern forms of communication continued to evolve and expand. Particularly, after the popularization of 5G, users' requirements have become rich and diverse, including calls, text messages, voice, games, videos, short videos, etc., and we call these multiple needs as \tr{multi-type data flow  (MDF)}. 
	%Clearly, each of requirements has its own characteristics. For example, the traditional communication method of telephone, which presents a stable demand, while the requirement for games or short videos  is rapidly changing. Therefore, traditional models are clearly not sufficient to describe modern communication needs.
	Different with classical teletraffic situation, MDF implies a queue problem for users with \textbf{random number of services} for a certain time.  \tr{This is because, for a person using a mobile phone, whether he is watching videos, playing games or making voice calls is completely random. Note that, the demand for these services varies, as does the durations (for example, voice calls are continuous demands, while games are ON-OFF type). Meanwhile, there may be an abrupt switch between demands, and multiple demands can also exist simultaneously. In addition, the device is unable to transmit the signal in continuous time, thus   MDF is also unable to be directly characterized by continuous time Markov chains (such as birth and death process). % According to the above reasons, traditional models are  not able to describe the MDF accurately.
	}
	
	\tr{There are some existing studies for   time-varying requirement system. \cite{ME-data} considered a special case that required data size follows a mixed-Erlang distribution and developed a numerical expression for the distribution of the steady-state queue length. \cite{servenumber} put the time-varying property on server number, which taking value $\lfloor s\rfloor$ or $\lfloor s\rfloor$+1.  \cite{OFDM}  studied time-varying states(active and silent) and set different requirements for different states. However, to the best of our knowledge, there is no model for general time-varying requirement case. Therefore, this paper establishes the 
		probability models and Erlang formulas for MDF in more general settings.} We denote the requirements of user $i$ at time $t$ as $X_i(t)$ (can be different for different $i$ and $t$).
	In Section \ref{section2}, we consider that the requirement  depends on the time of demand duration, and build   probability models  in three cases, non-tolerance, instant tolerance and delay. After assuming $X_i(t)$s are  \textit{i.i.d.} with respect to $i$ and $t$ with finite support set, Erlang formula is introduced in Section \ref{section3}. By discrete-time analysis, we find that  whatever $X_i(t)$s are variable or invariable with $t$, the the stable distribution of requirement of MDF and the discretization of EMLM (the discrete skeleton of continuous-time Markov chain) are the same. Therefore, we can solve the stable distribution of requirement for MDF by EMLM. 
	%A ON-OFF model is also discussed in Section \ref{section3}.
	The algorithm for pre-allocating  resources and examples are shown in Section \ref{section4}, and we conclude our results in Section \ref{section5}. %The proofs are put in \cite{wanzheng}. 
	The main contributions of this paper are summarized as follows.
	\begin{enumerate}
		\item For the case that requirements $X_i(t)$s depend on the time of demand duration, we  build probability models for MDF in three cases, non-tolerance, instant tolerance and delay, and obtain the blocking probability.
		
		\item \tr {When the  requirements of MDF are  identically distributed with respect to time, and service time is memoryless, we find the requirement process of MDF is    equivalent to   the discretization of EMLM. Consequently, we obtain the distribution for the total requirements of MDF. By error analysis, we find that  EMLM can be still used for MDF in this case.}

		\item The algorithm for pre-allocating resources is designed to guild the construction of base stations.
	\end{enumerate}
	
	Some notations are organized here. $X_i(t)$ denotes the requirement of user $i$ at time $t$, $A(t)$ denotes the  activated users at time $t$ and $|A(t)|$ denotes its number.  \tr{$\P$ always denotes the probability operator, and $\E$ denotes the expectation operator.}  We use $Poi(\lambda)$ and \tr{$Exp(\mu)$} to imply the random variables having Poisson distribution with rate $\lambda$ and exponential distribution with rate $\mu$ respectively, and $a \sim A$ means the random variable $a$ obeys the distribution $A$. We use $(\cdot)^k$ to represent the $k$ Cartesian product. $N$ always means the total number of users in a community while $C$ denotes the total number of resources. 	In the following of this paper, we consider a community with $N$ potential users for MDF.

	\section{Probability Models for Time-dependent	 MDF}\label{section2}
	Since the arrival of users for MDF are not significantly different with classical teletraffic problem, we still assume that the arrived number of users $I(t)$ follows a Poisson distribution with rate $\lambda N$, \textit{i.e.} $I(t)\sim Poi(\lambda N), \forall t$. In order to simplify the model, we assume that the serving time is discrete and  there is an uniformly bound for single use-time as $T$. This assumption also does not contradict with exponential distributed demand time in classical Erlang model since we can set the unit to be small enough to be close to continuous time, and  the probability of required time exceeding a large number $T$ is almost vanished. Note that if an user $i$ needs a $T_0(<T)$ demand duration, then we have $X_i(t)=0, \tau_i+T_0<t<\tau_i+T$, where $\tau_i$ is the demand start time of user $i$ and $t-\tau_i$ has uniformly distribution on $[T]:=\{0,1,\cdots, T\}$.

	It is apparently that the main different of different users are their start time $\tau_i$s, thus we assume that all activated users asking  requirement with a same distribution family $\{W(s), s\in\{0,1,2,\cdots, T\}\}$, \textit{i.e. } $X_i(t)\sim W(t-\tau_i),  i=1,2,\cdots, |A(t)|$.

	%Since the arrival and demand duration of users for MDF are not significantly different with classical teletraffic problem, we still assume that the arrived number of users follows a Poisson distribution with \textit{i.i.d.} exponential distributed required time.

	Given $C$ unit resources serving for a community,    
	we set standard to assess traffic overload. 
	\begin{enumerate}
		\item Non-tolerance. It implies that blockage occurs when demand exceeds total resources, \textit{i.e.}
		\begin{equation}\label{non-to_1}
		\frac{C}{\sum_{i\in A(t)}X_i(t)}\tr{\le} 1,\;\; \tr{t=0,1,2,\cdots.}
		\end{equation}
		
		\item Tolerance with threshold $\alpha\in(0,1)$. It means that a small amount of distortion is allowed. When demand exceeds resources, the data stream is compressed (such as reducing image or sound quality) to ensure smooth transmission. This is a very common method, and the maximum acceptable compression rate is set to be $1-\alpha$.  Thus the blockage happens when	
		\begin{equation}\label{to_1}
		\frac{C}{\sum_{i\in A(t)}X_i(t)}\tr{\le} \alpha,\;\; \tr{t=0,1,2,\cdots.}
		\end{equation}
		
		\item Loss as delay. \tr{Here we  consider  applications that are not delay-sensitive.} In this case, the un-transmitted information may be also put in the requirement in the next transmit-time, together with the new demands.  
	\end{enumerate}

	\subsection{Non-tolerance}
	Under the above settings, it is clear that 
	at time $t$, there are $I(t-T-1)$ users finishing their requirements, and $I(t)$ users joining in, thus when $t$ is large enough,
	%	\begin{equation}
	%	A(t)=A(t-1)+I(t)-I(t-T-1)\overset{d}{=}\begin{cases}
	%	\sum_{i=0}^t I(i), & t\le T;\\
	%	\sum_{i=0}^T I(i), & t> T,
	%	\end{cases}
	%	\end{equation}
	$
	|A(t)|=|A(t-1)|+I(t)-I(t-T-1)\overset{d}{=}
	\sum_{i=0}^T I(i),
	$
	where $I(t)\sim Poi(\lambda N)$, and $\overset{d}{=}$ means equal in distribution. Note that $I(t)$ is independent of $A(t-1)$ and $I(t-T)$,  thus the blocking probability is
	\begin{equation}\label{Non-tolerance}
	\begin{aligned}
	&\P\left(  \frac{C}{\sum_{i\in A(t)}X_i(t)}< 1\right)\\
	&=\sum_{k=1}^\infty \P(|A(t)|=k)\sum_{t^k-\vec{t}\in [T]^k}\frac{1}{(T+1)^k} \P\left( \sum_{i=1}^k W_i(t-t_i)>C\right), 
	\end{aligned}
	\end{equation}
	where  $W_i(t)\sim W(t)$ are independent, $t^k=(t,t,\cdots,t)_{1\times k}, \vec{t}=(t_1,t_2,\cdots,t_k)$.
	
	%	Note that by put $N$ in rate $\lambda N$, we assume that the potential users are infinity many. This is reasonable since $\P(|A(t)|\ge (1-\lambda)N)\approx 0$.
	%	
	
	By the property of Poisson process, $|A(t)|\sim Poi(\lambda N (T+1))$, thus
	\begin{equation}\label{expression of |A|}
	\begin{aligned}
	&\P\left(  \frac{C}{\sum_{i\in A(t)}X_i(t)}< 1\right)\\
	%&=\sum_{k=1}^\infty \P(|A(t)|=k)\sum_{t^k-\vec{t}\in [T]^k}\frac{1}{(T+1)^k} \P\left( \sum_{i=1}^k W_i(t-t_i)>C\right)\\
	&=\sum_{k=1}^\infty \frac{(\lambda N )^k}{k!}e^{-\lambda N(T+1)}\sum_{\vec{t}\in [T]^k} \P\left( \sum_{i=1}^k W_i(t_i)>C\right) .
	\end{aligned}
	\end{equation}
	
	Since the distributions of family $\{W(s), s\in\{0,1,2,\cdots, T\}\}$ are known (estimated by sampling for example), the blocking probability is obtained by \eqref{expression of |A|}.

	\subsection{Tolerance}
	It is obvious that tolerance case \eqref{to_1} is similar with non-tolerance \eqref{non-to_1} by replacing $C$ to $C/\alpha$, thus the blocking probability can be deduced directly that
	\begin{equation}\label{Tolerance}
	\begin{aligned}
	&\P\left(  \frac{C}{\sum_{i\in A(t)}X_i(t)}< \alpha\right)\\
	&=\sum_{k=1}^\infty \frac{(\lambda N )^k}{k!}e^{-\lambda N(T+1)}\sum_{\vec{t}\in [T]^k} \P\left( \sum_{i=1}^k W_i(t_i)>\frac{C}{\alpha}\right) .
	\end{aligned}
	\end{equation}

	\subsection{Delay}
	
	When the demand at a certain moment exceeds the amount of resources, the users will allocate the resources proportionally according to their  requirements, and the remaining demand will be added to the demand of the next unit of time, \textit{i.e.}, a (proportional) delay occurs. We denote the total requirements (arrival and delay) of user $i$ at time $t$ as $Y_i(t)$.
	%	We  ask that for   user $i$ in his activated time $[\tau_i, \tau_i+T]$, the wide that allocated for him satisfying
	%	\begin{equation}\label{tolerance for active model}
	%	\P\left( \frac{Cf}{\sum_{i\in A(t)}Y_i(t-\tau_i)}\le \alpha\right) \le 1-\varepsilon_1,\;\;\forall t\in[t_0, t_0+T],
	%	\end{equation}
	%	where $\varepsilon_1$ is a fixed small positive number. Since we consider the delay, 
	Clearly, $Y_i$ does not have the same distribution as $X_i$. But $Y_i$s are still \textit{i.i.d.}, in fact  $Y_i(\tau_i)=X_i(\tau_i)$, and when $\tau_i+1\le t\le \tau_i+T$,
	%	\begin{equation}
	%	\begin{aligned}
	%	&Y_i(t)-X_i(t)\\
	%	&=
	%	Y_i(t-1)\max\left\lbrace \left( 1- \frac{C}{\sum_{j\in A(t-1)} Y_j(t-1)}\right),0\right\rbrace, 
	%	\end{aligned}
	%	\end{equation}
	$
	Y_i(t)-X_i(t)=
	Y_i(t-1)\max\left\lbrace \left( 1- \frac{C}{\sum_{j\in A(t-1)} Y_j(t-1)}\right),0\right\rbrace. 
	$
	Replace $X_i$ by $Y_i$ in \eqref{Tolerance}, we can obtain the target probability for delay case.
	Note that $Y$ may not easily sampled as $X$, but if we denote $S(t)=\sum_{i\in A(t)}Y_i(t)$, we have
	\begin{equation}
	\begin{aligned}\label{S in delay}
	S(t)&=\max\{0, S(t-1)-C\}+\sum_{i\in A(t)} X_i(t)\\
	&=\max\{0, S(t-1)-C\}+\sum_{i\in A(t)} W_i(t-\tau_i)\\
	&\overset{\Delta}{=}S(t,\tau_1,\tau_2,\cdots,\tau_{|A(t)|}),
	\end{aligned}
	\end{equation}
	%	\begin{figure*}[hb] %hb代表放在文章底部，%ht为放在文章顶部 
	%		\centering
	%		\begin{align}\label{S in delay}
	%		&S(t)=\max\{0, S(t-1)-C\}+\sum_{i\in A(t)} X_i(t)\\
	%		&=\max\{0, S(t-1)-C\}+\sum_{i\in A(t)} X_i(\kappa_i)\\
	%		&\overset{\Delta}{=}S(t,\kappa_1,\kappa_2,\cdots,\kappa_{|A(t)|}),
	%		\end{align}
	%	\end{figure*}
	and the blocking probability becomes
	\begin{equation}\label{de}
	\begin{aligned}
	&\P\left( \frac{C}{\sum_{i\in A(t)}Y_i(t)}\le \alpha\right)=\P\left( \frac{C}{S(t)}\le \alpha\right)\\
	&=\sum_{k=1}^\infty \P(|A(t)|=k)\sum_{\vec{t}\in  [T]^k}\frac{1}{(T+1)^k} \P\left( S(t,\vec{\tau}=\vec{t})>\frac{C}{\alpha}\right)\\
	&=\sum_{k=1}^\infty \frac{(\lambda N )^k}{k!}e^{-\lambda N(T+1)}\sum_{\vec{t}\in  [T]^k}\P\left( S(t,\vec{\tau}=\vec{t})>\frac{C}{\alpha}\right),
	\end{aligned}
	\end{equation}
	\tr{where $\vec{t}=(t_1,t_2,\cdots,t_k), \vec{\tau}=(\tau_1,\tau_2,\cdots,\tau_k)$.}
	In order to estimate the distribution of $S(t)$, we assume that the requirements are  discrete (integers), which is applied in practice. Then the blocking probability can be write as \eqref{delay1}.
	\begin{figure*}[ht] %hb代表放在文章底部，%ht为放在文章顶部 
		\centering
		\begin{equation}\label{delay1}
		\begin{aligned}
		\P\left( \frac{C}{S(t)}\le \alpha\right)&=\sum_{k=1}^\infty \sum_{l=0}^k \P(I(t)=l)\P\left( \sum_{i=t-T}^{t-1}I(i)=k-l\right)C_k^l \sum_{\vec{t}\in([T]\setminus 0)^{k-l}} \frac{1}{T^{k-l}} \P(S(t,(0^l, t_1,t_2,\cdots, t_{k-l}))>\frac{C}{\alpha})\\
		%&=\sum_{k=1}^\infty \sum_{l=0}^k \P(I(t)=l)\P\left( \sum_{i=t-T}^{t-1}I(i)=k-l\right)C_k^l \sum_{\vec{t}\in([T]\setminus 0)^{k-l}} \frac{1}{T^{k-l}}\sum_{i=\lceil\frac{C}{\alpha  }\rceil}^P  \P(S(t,(0^l, t_1,t_2,\cdots, t_{k-l}))=i  )\\
		&=\sum_{k=1}^\infty \sum_{l=0}^k C_k^l e^{-\lambda N(T+1)}\frac{(\lambda N)^k}{l!(k-l)!}\sum_{\vec{t}\in([T]\setminus 0)^{k-l}} \sum_{i=\lceil\frac{C}{\alpha  }\rceil}^P  \P(S(t,(0^l, t_1,t_2,\cdots, t_{k-l}))=i  ).
		\end{aligned}
		\end{equation}
	\end{figure*}
	The only unknown term is $\P(S(t,(0^l, t_1,t_2,\cdots, t_{m})=i) )$, it may obtained by recursion as   shown in \eqref{delay2}.
	\begin{figure*}[hb] %hb代表放在文章底部，%ht为放在文章顶部 
		\centering
		\begin{equation}\label{delay2}
		\begin{aligned}
		&\P(S(t,(0^l, t_1,t_2,\cdots, t_{m})=i) )\\
		&=\sum_{j=0}^\infty \P(I(t-T-1)=j) \sum_{a=0}^P \P(S(t-1, (T^j, t_1-1, t_2-1\cdots, t_{m}-1))=a )\\
		&\;\;\;\;\cdot  \P\left( S(t, (0^l,t_1, t_2\cdots, t_{m}) )=i |S(t-1, (T^j, t_1-1, t_2-1\cdots, t_{m }-1))=a \right) \\
		%&=\sum_{j=0}^\infty \P(I(t-T-1)=j) \sum_{a=0}^P \P(S(t-1, (T^j, t_1-1, t_2-1\cdots, t_{m}-1))=a ) \P\left(  \sum_{k=1}^l W_k(0)+\sum_{k=1}^m W_k(t_k)=i -\max\{0, a -C\}\right)\\
		&=\sum_{j=0}^\infty e^{-\lambda N}\frac{(\lambda N)^j}{j!} \sum_{a=0}^P \P(S(t-1, (T^j, t_1-1, t_2-1\cdots, t_{m}-1))=a )  \P\left(  \sum_{k=1}^l W_k(0)+\sum_{k=1}^m W_k(t_k)=i -\max\{0, a -C\}\right)
		\end{aligned}
		\end{equation}
	\end{figure*}
	The calculation of $\P\left(  \sum_{k=1}^l W_k(0)+\sum_{k=1}^m W_k(t_k)=i -\max\{0, a -C\}\right)$ is just the problem with form $\P(\sum_{i=1}^m W_i(t_i)=j )$ which is the same in the previous section, and the initial value  is 
	$
	\P(S(0,(0^m))=j )=\P\left( \sum_{i=1}^m W_i(0)=j \right),
	$
	which is also obtained by the distribution of $W$.

	\section{Erlang Formula for MDF}\label{section3}
	In Section \ref{section2}, we consider the case that the requirement depends on time $t$ and set the probability models to calculate the blocking probabilities. However, the large number of convolution operations makes the complexity of the algorithm very high. In this section, we further assume that if  user $i$ is activated at time $s$ and $t$, the requirements $X_i(s)$ and $X_i(t)$ are identically distributed, \textit{i.e.} the distribution of requirement does not depend on time. It is reasonable  since the demands of data stream are diverse and varied with respect to time in MDF.

	To be realistic, we still assume the requirement is discrete as in the references \cite{BeyondEr}, \cite{tolerance2012},	\cite{THandTolerance}. Specifically, if user $i$ is activated at time $t$, then $X_i(t)\in\{b_1,b_2,\cdots,b_K\}\subset \mathbb{N}^+$ with distribution
	\begin{equation}\label{X}
	\P(X_i(t)=b_k)=a_k,\;\;k=1,2,\cdots,K.
	\end{equation} 
	
	\tr{In addition, we use $t_s$ to denote the unit time and 
		consider a typical case that a user activated at time $at_s$ is still activated at time $(a+1)t_s$ with probability $p$ and leave with probability $1-p$, where $p\in(0,1)$ (only depends on $t_s$,  in fact $p\to 1$ when $t_s \to 0$, since if $t_s$ is small, any user must have more then 1 unit required time). This assumption is the memoryless property of the required time, which is widely set up in the queuing problem \cite{queue1}.} 
	
	For arrival users, we suppose that the numbers of users arriving at time $t=0,t_s,2t_s,\cdots$ all follow a Poisson distribution with parameter $\lambda t_s$.
	\tr{ 
		\begin{definition}
			Under the above setting, we call the community as a $MDF(\lambda,p,t_s)$ system, and the total requirement at time $t$ is denoted as $S_{\lambda, p, t_s}(t)$.
	\end{definition}}
	
	% Another different between our model in Section \ref{section2} and Erlang model is the service time. 
	% 
	% 
	% 
	% Recall that in traditional queue problem,  required time of users are set to be negative exponential distributed with parameter $\mu$. Since we can set $T$ be large enough, the different between bounded time in our model and unbounded time in Erlang model can be ignored, and then the core different is the discrete service time in our model and the continuous service time in Erlang model. 
	\tr{ 
		Note that this setting is covered by our model in Section\ref{section2} once we take $T$ large enough and set $X_i(t)=0$ if user $i$ has already left at time $t$.} 
	
	\tr{  In the remaining of this section,  
		we only consider the ``tolerance" case, since it is a common event in MDF and covers ``non-tolerance".  We first introduce the existing Erlang model and show the different of settings between the Erlang model and the above case. Then we exhibit a mathematical analysis to overcome these differences. }

	\subsection{Erlang Multirate Loss Model}
	In existing models, \tr{time $t$ is a continuous parameter, and requirement for an activated user is fixed}. Specifically,  it is considered that if  user $i$ is activated at time $s$ and $t$, then 
	\begin{equation}\label{assume of EMLM}
	X_i(s)=X_i(t),
	\end{equation}
	and \tr{required time of users are set to be negative exponential distributed.}
	Following the above settings, the total  requirement in a community $S_{EMLM}(t)=\sum_{i\in A(t)}X_i(t)$ at time $t$ is a stationary birth and death process. Certainly, if
	users have $\lambda$ arrival rate with $Exp(\mu)$ distributed required time, then the distribution of $S_{EMLM}$ satisfies the equations
	\begin{equation}\label{EMLM}
	\sum_{k=1}^K \frac{\lambda_k}{\mu}\left( b_kq(j-b_k)\right) =jq(j), \;\; j\in (\mathbb{N}^K\cdot \{b_1,b_2,\cdots, b_K\}),
	\end{equation}
	where $q(j)=\P(S_{EMLM}=j), \lambda_k=\lambda N a_k$ and
	$
	(\mathbb{N}^K\cdot \{b_1,b_2,\cdots, b_K\})=\{ s| s=\sum_{i=1}^K d_i b_i, d_i\in \mathbb{N}, i=1,2,\cdots,K \}.
	$
	Obviously, $\P(j>N\max_k\{b_k\})=0$, thus
	\begin{equation}\label{constrain_EMLM}
	\sum_{j\in (\mathbb{N}^K\cdot \{b_1,b_2,\cdots, b_K\}), j\le N\max_k\{b_k\}} q(j)=1,
	\end{equation}
	equations \eqref{EMLM} and \eqref{constrain_EMLM} can be solved iteratively, and the blocking probability (for tolerance) is
	\begin{equation}\label{gs}
	\P(S_{EMLM}>C/\alpha)=\sum_{j\in (\mathbb{N}^K\cdot \{b_1,b_2,\cdots, b_K\}), \frac{C}{\alpha}<j\le N\max_k\{b_k\}} q(j).
	\end{equation}
	\tr{
		\begin{remark}\label{re1}
			It is worthy to note that if \eqref{assume of EMLM} is changed to be
			\begin{equation}\label{assume of EMLM2}
			X_i(s)\overset{d}{=}X_i(t),
			\end{equation}
			where $\overset{d}{=}$ means equal in distribution,
			the result still holds. It is easy to prove and our analysis in next part can also implies this statement, as shown in Corollary \ref{co}.
		\end{remark}
	}

	\subsection{Distribution analysis for $S_{\lambda, p,t_s}$}
	Without assumption \eqref{assume of EMLM} and exponential distributed service time, we can not use EMLM for MDF directly. However, we still consider that the total demand process is stationary. \tr{In order to analyze the distribution of $S_{\lambda, p,t_s}$ for MDF, we find the discrete skeleton of EMLM process.}
	%transform the continuous Markov process of EMLM to make it close to the process of MDF.} Specifically, 
	%we discretize the time and  divide the total requirement into two parts, the demand that continues and the new demand that arrives.
	
	\begin{enumerate}
		\item Discretization. We discretize the time, which is a common approach for continuous Markov process. Specifically, we treat all arriving time $t\in [st_s, (s+1)t_s)$ as $st_s$ and all leaving time $t\in [st_s, (s+1)t_s)$ as $(s+1)t_s$, where $t_s$ is a (small enough) unit time. Under this discretization, the relative total requirement $\hat{S}$ satisfies 
		\begin{equation}\label{error}
		\begin{aligned}
		\varepsilon_0(t)&:=|\hat{S}(t)-S_{EMLM}(t)|\\
		&\le (\max_k b_k)(P_0(t_s)+\sum_{i\in A(t)}1_{\{l_i<t_s\}}),\;\;\forall t\ge 0,
		\end{aligned}
		\end{equation}
		where $P_0(t_s)\sim Poi(\lambda t_s N), l_i\sim Exp(\mu)$ are the number of users arriving and leaving in a time interval of length $t_s$. Obviously,
		$
		\E \varepsilon_0 \le  N(\max_k b_k)(\lambda t_s+\P(Exp(\mu)<t_s)),
		$
		thus 
		$\lim_{t_s\to 0}  \varepsilon_0 \to 0$, \textit{a.s.}
		%and we can use the distribution of $\hat{S}$ to approximate $S$.
		
		%		 $\lim_{t_s\to 0}\varepsilon_0\overset{\P}{\to}0$,

		\item Continuous requirement. After discretization, any user activating at time $(t-1)t_s$ flips a coin with $\P(Y=1)=\int_{t_s}^\infty \mu e^{-\mu x}dx:=\tr{p_\mu}$ and $\P(Y=0)=\tr{1-p_\mu}$ to decided leave or not, thus the continuous requirement at time $tt_s$ is
		$
		\hat{S}_c(tt_s)=\sum_{i=1}^{A((t-1)t_s)} X_iY_i. 
		$
		By traditional Erlang formula it is easy to deduce the distribution of $A(t)$ when $N$ is large enough,
		$
		\P(A(t)=n)=\frac{\rho^n/n!}{\sum_{i=0}^N \rho^i/i!}\approx \frac{\rho^n/n!}{\sum_{i=0}^\infty \rho^i/i!},\;\;\rho=\lambda t_s N/\mu,
		$
		\textit{i.e.} $A(t)$ can be estimated as a Poisson process with rate $\lambda t_s N/\mu$. In addition, since $Z_i:=X_iY_i$ has distribution
		$
		\P(Z_i=0)=1-\tr{p_\mu},\;\; \P(Z_i=b_k)=\tr{p_\mu} a_k,
		$
		therefore, $Z_i$ can also considered as a requirement random variable and by \cite[Theorem 3.7.4]{martingle},
		$
		\hat{S}_c(tt_s)=\sum_{k=1}^K b_k P^1_k,
		$
		where $P^1_k\sim Poi(\frac{\lambda \tr{p_\mu} t_s a_k N}{\mu})$.
		
		\item Arrival requirement. The number of arrival users in $[tt_s, (t+1)t_s)$ is $I\sim Poi(\lambda t_s N)$ and according to \eqref{X}, the number of packets  arriving in $[tt_s, (t+1)t_s)$ is 
		$
		\hat{S}_a(tt_s)=\sum_{k=1}^K b_kP_k
		$	
		where 
		$
		P_k\sim Poi(\lambda t_s a_k N).
		$ 
	\end{enumerate}
	In conclusion, we have the distribution of $\hat{S}$ as
	\begin{equation}\label{hat S}
	\hat{S}(t)=\hat{S}_c(tt_s)+\hat{S}_a(tt_s)=\sum_{k=1}^K b_k P^2_k, t\in[tt_s, (t+1)t_s),
	\end{equation}
	where $P^2_k\sim Poi(\lambda t_s a_kN(1+\tr{p_\mu}/\mu))$.
	
	\tr{It is notable that the discretization $\hat{S}$ is exactly the total requirement $S_{\lambda,p_\mu, t_s}$, therefore, we has the following result.
		\begin{theorem}\label{tt1}
			The distribution of $S_{\lambda,p,t_s}$ is
			\begin{equation}\label{dis}
			S_{\lambda,p,t_s}=\sum_{k=1}^K b_k P^2_k,
			\end{equation}
			where $P^2_k\sim Poi(\lambda t_s a_kN(1-pt_s/\ln p))$ are independent. Furthermore, if $\lim_{t_s\to 0} -\ln p/t_s=\mu$, then $S_{\lambda,p,t_s}$  is close to the $S_{EMLM}$ with arrival rate $\lambda$ and $Exp(\mu)$ distributed required time, \textit{i.e.}
			\begin{equation}\label{===}
			\lim_{t_s\to 0}|S_{\lambda,p,t_s}-S_{EMLM}|=0, \textit{a.s.}.
			\end{equation}
			%where $\overset{\P}{\to}$ means convergence under probability.
		\end{theorem} 
		Since whether $\eqref{assume of EMLM}$ or $\eqref{assume of EMLM2}$ holds, \eqref{===} is both satisfied,  this implies the truth of statement in Remark \ref{re1}, which is concluded in the following corollary.
		%\begin{corollary}\label{co}
		%	 	For a community with $N$ people, suppose the arrived number of users follows a Poisson distribution with rate $\lambda$, and  required time of users is negative exponential distributed with parameter $\mu$.  If the requirement of any activated user at any time $t$ obeys the distribution
		%$
		%	 \P(X_i(t)=b_k)=a_k,\;\;k=1,2,\cdots,K,
		%$
		%	 then the blocking probability for tolerance rate $\alpha$ and base resources $C$ is 
		%	 \begin{equation}\label{co1}
		%	 \P(S>C/\alpha)=\sum_{j\in (\mathbb{N}^K\cdot \{b_1,b_2,\cdots, b_K\}), \frac{C}{\alpha}<j\le N\max_k\{b_k\}} q(j),
		%	 \end{equation}
		%	 where $q$ can be solved by \eqref{EMLM} and \eqref{constrain_EMLM}.
		%\end{corollary}
		\begin{corollary}\label{co}
			The blocking probability remains the same when \eqref{assume of EMLM} is replaced by \eqref{assume of EMLM2}.
		\end{corollary}
	}
	\subsection{Blocking Probability for $MDF(\lambda,p,t_s)$}
	\tr{	Following the  above deduction, we can use the Blocking Probability of EMLM to estimate the one of  $MDF(\lambda,p,t_s)$, and the estimation error tends to 0 as shown in the following.
		\begin{theorem}\label{t1}
			Consider a $MDF(\lambda,p,t_s)$ system with $N$ people, and suppose the requirement of any activated user at any time $t$ obeys the distribution
			$
			\P(X_i(t)=b_k)=a_k,\;\;k=1,2,\cdots,K.
			$
			If $\lim_{t_s\to 0} -\ln p/t_s=\mu$,
			then the blocking probability for tolerance rate $\alpha$ and base resources $C$ satisfies that as  $ t_s\to 0$, the term
			\begin{equation}\label{bp2}
			\begin{aligned}
			\delta\overset{\Delta}{=} &\bigg|\P(S_{\lambda,p,t_s}>C/\alpha)-\sum_{j\in (\mathbb{N}^K\cdot \{b_1,b_2,\cdots, b_K\}), \frac{C}{\alpha}<j\le N\max_k\{b_k\}} q(j)\bigg|\\
			%&\overset{\textit{a.s.}}{\to}0,
			\end{aligned}
			\end{equation}
			tend to $0$ almost sure, where $q$ can be solved by \eqref{EMLM} and \eqref{constrain_EMLM} with parameter $(\lambda,\mu)$.
		\end{theorem}
		%	\pf
		%See \cite{wanzheng} for details.
		%	\e
		\pf Without loss of generality, we assume $p=e^{-\mu t_s}$.
		By \eqref{error}, we have $\forall x\ge 0,$
		$
		\P(S_{EMLM}>x+\varepsilon_0)\le \P(S_{\lambda,p,t_s}>x)\le  \P(S_{EMLM}>x-\varepsilon_0).
		$
		Thus
		$
		\delta\le \sum_{j\in (\mathbb{N}^K\cdot \{b_1,b_2,\cdots, b_K\}), \frac{C}{\alpha}-\varepsilon_0<j\le \frac{C}{\alpha}+\varepsilon_0} q(j)\le 2\varepsilon_0.
		$
		%			\begin{equation}
		%			\begin{aligned}
		%				%&\bigg|\P(S_{\lambda,p,t_s}>C/\alpha)-\sum_{j\in (\mathbb{N}^K\cdot \{b_1,b_2,\cdots, b_K\}), \frac{C}{\alpha}<j\le N\max_k\{b_k\}} q(j)\bigg|\\
		%				\delta&\le \sum_{j\in (\mathbb{N}^K\cdot \{b_1,b_2,\cdots, b_K\}), \frac{C}{\alpha}-\varepsilon_0<j\le \frac{C}{\alpha}+\varepsilon_0} q(j)\le 2\varepsilon_0.
		%			\end{aligned}
		%	\end{equation}
		Since $\varepsilon_0\overset{a.s.}{\to}0,$ as $t_s\to0$, the proof  completes.
		\e
	}
	
	\tr{
		Theorem \ref{t1} implies that under the setting of this section,  once the unit time $t_s$ is small enough, we can use the blocking probability of EMLM to estimate the one of $MDF(\lambda,p,t_s)$ system, without large number of calculations for convolution.   
	}	

	\section{Simulation}\label{section4}
	%In this section, an algorithm for construction of base resources is displayed and we also show the accuracy of our model in Section \ref{section2} by a toy example.

	\subsection{The Algorithm for Pre-allocating  Resources}
	Firstly we assume that $t_s$ is small enough in order to use Theorem \ref{t1}, and parameters $(N, \lambda, p, K, a_1,a_2,\cdots,a_K,b_1,b_2,\cdots,b_K,\sigma)$ are all known,   which can be estimated from a large number of samples. Then given a tolerance rate $\alpha$, or the maximum  distortion rate $1-\alpha$, and the acceptable blocking probability $\varepsilon$, the minimum $C$ is the output of Algorithm \ref{al}.
	\begin{algorithm}[t]\label{al}
		\caption{Algorithm for Pre-allocation}
		\label{alg:AOA}
		\renewcommand{\algorithmicrequire}{\textbf{Input:}}
		\renewcommand{\algorithmicensure}{\textbf{Output:}}
		\begin{algorithmic}[1]
			\REQUIRE $(N, \lambda, p, K, a_1,a_2,\cdots,a_K,b_1,b_2,\cdots,b_K,\sigma), \alpha, \varepsilon, t_s$
			\ENSURE $C$
			
			\STATE $q(0)=1$, \tr{$\mu=-\ln p/t_s$,} $C=\max_k b_k$.
			\STATE calculate $q(j)$ by \eqref{EMLM}  recursively.
			\STATE $q(j)=q(j)/\sum_{i\in (\mathbb{N}^K\cdot \{b_1,b_2,\cdots, b_K\}), i\le N\max_k\{b_k\}} q(i)$.
			\WHILE {1} 
			\STATE calculate $\P(S>C/\alpha)$ by \eqref{gs}.
			\IF {$\P(S>C/\alpha)>\varepsilon$}
			\STATE $C=C+1$.
			\ELSE
			\STATE break.
			\ENDIF
			\ENDWHILE
			
			\RETURN $C$.
		\end{algorithmic}
	\end{algorithm}
	
	\subsection{A Toy Example}
	\tr{In this part, we simulated the blocking probabilities of EMLM and $MDF(\lambda,p,t_s)$ system to show the correction of Theorem \ref{tt1} and \ref{t1}. }
	Specifically, we choose $N=10000, \lambda=0.001, \mu=0.5, t_s=0.0001, p=e^{-\mu t_s}\approx 0.99995$ and $\sigma=0.5$. The requirement $X$ obeys the distribution
	$
	X_1\sim \begin{pmatrix}
	1&2&4&8\\
	0.45&0.35&0.15&0.5
	\end{pmatrix},
	$
	\tr{and
		$
		X_2\sim \begin{pmatrix}
		1&3&5&7&9&11\\
		0.15&0.1&0.3&0.25&0.15&0.05
		\end{pmatrix}.
		$}
	\tr{By replacing step 5 in Algorithm \ref{al} to related calculations and  fixing parameters, we can draw the $\varepsilon-C$ curve according to \eqref{EMLM} and \eqref{dis}. On the other hand, based on the setting in Section \ref{section3}, we can also use   Monte-Carlo method to simulate the blocking probability of the MDF system, as shown in Figure \ref{F1}. 
		\begin{figure}[t]
			\centering
			\includegraphics[width=0.38\textwidth]{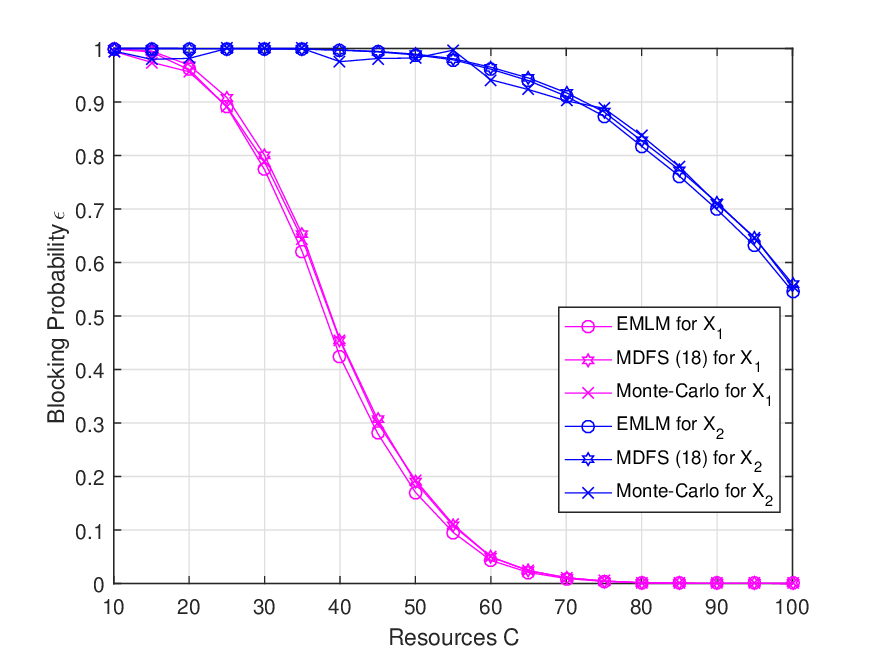}
			\caption{\tr{Blocking Probability Curve}}
			\label{F1}
		\end{figure}
		Apparently, Under both two  requirements, the three curves are very close,   which implies the accuracy of our analysis.  Unfortunately, we don't have enough data to analyze how our probabilistic model differs from reality (It is also a reason we build our model since there is no enough data to guild the construction of base station directly). However, Poisson arrival and  memoryless service time are long-standing assumptions about teletraffic problems, and under these assumptions, our mathematical analysis illustrates the accuracy of our models.}

	\section{Conclusion}\label{section5}
	In this paper, the probabilistic models for the  MDF are established. For general case, we analyzes blocking probabilities of MDF in three states: non-tolerance, tolerance and delay, as shown in \eqref{expression of |A|}, \eqref{Tolerance} and \eqref{delay1}. When the requirements are co-distributed with respect to time \tr{and the service time is memoryless, a mathematical proof shows that EMLM  is applicable to MDF, as stated in Theorem \ref{t1}.} An method to pre-allocate  resources for communication society is given in Algorithm \ref{al} and a toy example implies the correction of our theoretical deductions.

\end{document}